\RequirePackage{filecontentsdef}
\documentclass[conference]{IEEEtran}
\IEEEoverridecommandlockouts

\usepackage{cite} 
\usepackage{amsmath,amssymb,amsfonts} 
\usepackage{algorithmic} 
\usepackage{graphicx} 
\usepackage{textcomp} 
\usepackage{xcolor} 
\usepackage{hyperref} 
\usepackage{pgfplots}
\usepackage{filecontents}

\pgfplotsset{compat=1.18}

\newcommand{\rrp}{{\mathbb{R}^{p \times p}}} 
\newcommand{\cc}{\mathcal{C}} 
\newcommand{\gggg}{\mathcal{G}} 
\newcommand{\vvv}{\mathcal{V}} 
\newcommand{\eee}{\mathcal{E}} 
\newcommand{\cb}{\mathbf{C}} 
\newcommand{\mathbfit}[1]{\textbf{\textit{#1}}} 
\newcommand{\mat}[1]{\mathbf{#1}}

\begin{document}

\title{Generating Correlation Matrices with Graph Structures Using Convex Optimization
\thanks{This work was supported by the Agence Nationale de la Recherche under
the France 2030 programme, reference ANR-23-IACL-0006.}
}

\author{\IEEEauthorblockN{Ali Fahkar\IEEEauthorrefmark{1}, Kévin Polisano\IEEEauthorrefmark{1}, Irène Gannaz\IEEEauthorrefmark{3}, Sophie Achard\IEEEauthorrefmark{1}} \\
\IEEEauthorblockA{
\IEEEauthorrefmark{1} Univ. Grenoble Alpes, CNRS, Grenoble INP, Inria, LJK, F-38000 Grenoble, France \\
\IEEEauthorrefmark{3} Univ. Grenoble Alpes, CNRS, Grenoble INP, G-SCOP, 38000 Grenoble, France }
}

\maketitle

 \begin{abstract}
   	This work deals with the generation of theoretical correlation matrices with specific sparsity patterns, associated to graph structures. We present a novel approach based on convex optimization, offering greater flexibility compared to existing techniques, notably by controlling the mean of the entry distribution in the generated correlation matrices. This allows for the generation of correlation matrices that better represent realistic data and can be used to benchmark statistical methods for graph inference.
\end{abstract}

	
\section{Introduction} \label{sec:intro}

Graphical models provide a way to represent dependencies between random variables. This field has seen a lot of interests in recent years, see e.g.~\cite{kimmig2015lifted, engel2017overview, pearl2014probabilistic, koller2009probabilistic}, \cite[Chapter 7]{Giraud} and references therein. The range of applications is broad, including genetics~\cite{grechkin2015pathway}, proteins study~\cite{akbani2014pan}, disease characterization~\cite{armour2017network}, functional brain connectivity~\cite{huang2010learning}, or risk management~\cite{hull2012risk}. The principle is to infer a graph structure associated with the correlation matrix or the precision matrix (inverse of the correlation matrix). To assess the quality of estimation procedures, simulation studies are essential. This requires generating (theoretical) correlation or precision matrices associated to a given graph structure which involves imposing zeros, that is often non-trivial. The aim of this paper is to present a method for generating such matrices.

Among the generation procedures proposed in the literature for correlation matrices, we can cite the so-called \textit{vines} and \textit{onion} procedures~\cite{lewandowski2009generating}, based on the Beta distribution established by \cite{joe2006generating}. Alternatively, \cite{pourahmadi2015distribution, cordoba2020generating} use the Cholesky decomposition to generate correlation matrices. We refer to these articles for a bibliographic overview of existing methods. In~\cite{achard2022generation}, the distribution of brain connectivity correlations was found to be centered around positive values. However, the proposed methods all generate correlation matrices whose entry distribution is centered around zero. Our objective is to propose a new approach based on convex optimization that allows control over this distribution, particularly its mean.

This paper is organized as follows. Section~\ref{sec:notation} introduces key definitions and notations. Section~\ref{sec:related} reviews related work. Section~\ref{sec:proposed} describes the proposed approach, and  Section~\ref{sec:results} presents the results and comparison with other approaches.

\section{notation} \label{sec:notation}

For a fixed dimension $p$, we denote matrices in $\rrp$. A real symmetric $p \times p$ matrix $\mathbf{A}$ is positive semidefinite (PSD) if $\mathbfit{x}^\top \mathbf{A} \mathbfit{x} \geq 0$ for all $\mathbfit{x} \in \mathbb{R}^p$. It is positive definite (PD) if $\mathbfit{x}^\top \mathbf{A} \mathbfit{x} > 0$ for all non-zero $\mathbfit{x} \in \mathbb{R}^p$.

Let $\mathbfit{x} \in \mathbb{R}^{p}$ be a random vector with covariance matrix $\mathbf{\Sigma}$, defined as:
$
	\mathbf{\Sigma} = \mathbb{E} [(\mathbfit{x} - \mathbb{E}[\mathbfit x])(\mathbfit{x} - \mathbb{E}[\mathbfit x])^\top] = (\sigma_{i,j})_{i,j=1,\dots,p}.
$
The corresponding correlation matrix $\cb \in \mathcal{C}$ is defined by $
c_{ij} = \frac{\sigma_{ij}}{\sqrt{\sigma_{ii} \sigma_{jj}}}$ for all $(i,j)\in\lbrace 1,\dots,p\rbrace$, or equivalently 
$\cb = (\text{diag}(\mathbf \Sigma))^{-\frac{1}{2}} \ \mathbf \Sigma \ (\text{diag}(\mathbf \Sigma))^{-\frac{1}{2}} $.
The set $\mathcal{C}$ of correlation matrices satisfies
\begin{equation}
	\label{eq:condition}
	\forall\cb \in \mathcal{C} \quad \text{diag}(\cb) = 1, \
	\forall i, j \in \{1, 2, \dots, p\} \ -1\leq c_{ij} \leq 1.
\end{equation}
From a generative perspective, working with correlation matrices and precision matrices is similar. Therefore, we focus on correlation matrices. Generating a correlation matrix involves constructing a symmetric PSD matrix that satisfies condition~\eqref{eq:condition}~\cite[Problem 7.1.]{horn2012matrix}.

A graph $\gggg$ is a mathematical structure used to represent pairwise relations between objects. Formally, a graph is defined as  $\gggg = (\vvv, \eee)$,
where $\vvv= \{v_1, v_2, \dots , v_p\}$ is the set of vertices, and $\eee \subset \vvv \times \vvv$ is the set of edges. An edge is a pair of vertices connected by the graph. We look for a correlation 
matrix $\cb$ that is associated with the graph, meaning it satisfies $c_{ij} = 0$ if $(i, j) \notin \eee$ and $c_{ij} \neq 0$ otherwise, for all~$(i,j)\in\vvv \times \vvv$ with $i\neq j$. The weights of the edges in the graph correspond to the values in the correlation matrix. 
In other words, we impose sparsity on certain relationships. We define $\overline{\eee}$ as the set of non-edges, corresponding to the zero entries in the matrix $\cb$. Thus, our objective is to generate a correlation matrix with a prescribed set $\overline{\eee}$ of zero entries. This problem can be viewed as a matrix completion problem~\cite[Chapter 10]{vandenberghe2015chordal}. 

In the following, we denote by $\cc(\gggg)$ the set of correlation matrices associated with a given graph $\gggg$, that is satisfying the following constraints: 
\begin{equation}
	\label{eq:constraint1}
	\cb = (c_{ij})  \text{ PSD and satisfies \eqref{eq:condition}}, \quad c_{ij} = 0, \ (i, j) \notin \eee.
\end{equation}
We consider different graph structures:
\begin{itemize}
    \item \emph{Erdős-Rényi random graphs}~\cite{erdos1960evolution}, probabilistic graphs in which edges between nodes are formed independently with a fixed probability;
    \item \emph{Barabási-Albert random graphs}~\cite{albert1999diameter}, scale-free networks that generate graphs using preferential attachment, where new nodes are more likely to connect to already well-connected nodes;
    \item \emph{Watts-Strogatz random graphs}~\cite{watts1998collective}, small-world network models that combine high clustering coefficient with short average path lengths;
    \item \emph{Stochastic Block Models}~\cite{abbe2018community}, generative models for networks where nodes are partitioned into blocks, with different probabilities of connections within and between blocks;
    \item \emph{Chordal graphs}, graphs in which every cycle of length greater than three has a chord (an edge connecting two non-adjacent vertices in the cycle).
\end{itemize}
A key characteristic of the graph structure is graph density, which is the ratio of the number of edges and the number of possible edges. For a graph with $|\eee|$ edges and $p$ vertices, the graph density is defined as \mbox{$d = \frac{2|\eee|}{p(p-1)}$}. 

In our study, we generate a chordal graph by starting with a Barabási-Albert graph and adding edges as needed to satisfy the chordal graph properties. 
Node numbering is the process of assigning a unique integer to each node in a graph~\cite[Chapter 2]{vandenberghe2015chordal}. While node numbering can be arbitrary, \emph{Perfect Elimination Ordering} (PEO) is an ordering of the vertices $v_1, v_2, \dots, v_n$ such that for any vertex $v_i$, the set of neighbors \mbox{$\{v_j \mid j > i \text{ and } (v_i, v_j) \in \eee\}$} form a \emph{clique}, that is a subgraph in which every pair of distinct vertices is adjacent.
A graph has a PEO if and only if it is chordal~\cite[Chapter 4]{vandenberghe2015chordal}. 

\section{Related Work} \label{sec:related}

The generation of a (theoretical) correlation matrix has been proposed for example in~\cite{joe2006generating} with a characterization of the uniform distribution over the space of correlation matrices. Common approaches include the \textit{vines} and \textit{onion}~\cite{lewandowski2009generating}. For a broader review of available techniques, we refer to the bibliographic surveys in~\cite{pourahmadi2015distribution, cordoba2020unifying}. Most existing methods, however, cannot be extended to generate correlation matrices under structural constraints. In particular, they do not allow for the generation of correlation matrices that are associated to a given graph $\gggg$. Below, we present the methods that, to our knowledge, can generate such matrices.

\subsection{Chordal Graphs and Cholesky Decomposition} \label{subsec:chordal_graphs}
The first approach relies on the Cholesky decomposition.
Let $\mathcal{U}\in \rrp$ be the set of upper triangular matrices, with positive diagonal entries and rows normalized to~1. Define $\mathcal{U}(\gggg)\subset \mathcal{U}$ as the subset where $u_{ij}=0$ for all \mbox{$(i,j)\in \overline{\eee}$.} If the graph is ordered (PEO), it is possible to generate the Cholesky factor $\mat{U}$ in $\mathcal{U}(\gggg)$ by imposing $u_{ij}=0$ for all $(i, j)\in \overline{\eee}$ to obtain a matrix $\cb=\mat{U}\mat{U}^\top\in \cc(\gggg)$ such that ${m}_{ij}=0$ for all $(i,j)\notin \eee$. For further details, we refer to~\cite{cordoba2020generating}.
In~\cite{pourahmadi2015distribution}, the authors propose a polar writing of the entries of the Cholesky factor $\mat{U}\in\mathcal{U}$, and establish the probability distribution of the induced quantities such that the resulting distribution is uniform over the set $\cc$.
Using this polar parametrization, it is straightforward to incorporate the constraint $u_{ij}=0$ for all $(i,j)\in \overline{\eee}$, to obtain $\mat{U}\in\mathcal{U}(\gggg)$. In~\cite{cordoba2020generating} the former constraint is already taken into account. The proposed generation method is based on the Metropolis-Hastings algorithm~\cite{chib1995understanding} and yields a uniform distribution over $\cc(\gggg)$.

As mentioned above, these methods assume that the graphs are ordered, which is generally not the case. In fact, only chordal graphs can be perfectly ordered~\cite[Chapter 4]{vandenberghe2015chordal}. Therefore, this approach is only available for chordal graphs.

\subsection{Diagonal Dominance and Partial Orthogonalization} \label{subsec:all_graphs}

To the best of our knowledge, two methods have been proposed in the literature to generate correlation matrices associated to a given graph $\gggg$ without requiring a chordal structure: diagonal dominance and partial orthogonalization.

Diagonal dominance as proposed in~\cite{cordoba2020unifying} can be used to generate a PD matrix. The idea is to construct a symmetric matrix $\tilde{\cb}$, where the elements $\tilde{c}_{ij}$ are chosen uniformly at random from the interval $[-1, 1]$ if $(i, j) \notin \bar{\eee}$, and $\tilde{c}_{ij} = 0$ otherwise. Then, using the following update rule: 
\begin{footnotesize}
	\begin{equation}\label{eq:rule}
	\forall i \in \{1, \dots, p\} \quad \text{do } \tilde{c}_{ii} \leftarrow \sum_{\substack{j = 1, \dots, p \\i\neq j}}\lvert{\tilde{c}_{ij}}\rvert + \text{random positive perturbation}.
    \end{equation}
\end{footnotesize}
This follows from the Gershgorin theorem~\cite{salas1999gershgorin}. In Eq. \eqref{eq:rule}, if the random positive perturbation is omitted, the method instead produces a PSD matrix. Then, define \mbox{$\cb=\text{diag}({\tilde{\cb}})^{-1/2}\, \tilde{\cb}\, \text{diag}(\tilde{\cb})^{-1/2}$} to recover a matrix in $\cc$. However, a major drawback of this approach is that it yields correlation matrices with very low off-diagonal values.

The partial orthogonalization method, proposed in~\cite{cordoba2020generating}, provides an alternative that works for non-chordal graphs. The idea is to start with an initial matrix $\cb$ whose zero entries are included in the desired sparsity pattern $\overline{\eee}$. The additional edges of $\cb$ are then removed using a modified Gram-Schmidt based partial orthogonalization process. The idea is to write $\cb$ as $\cb=\mathbf{Q}\mathbf{Q}^\top$ and then iteratively orthogonalizes every row
$\mathbfit{q}_{i.}$ with respect to the set of rows $\lbrace{\mathbfit{q}_{j.} s.t. (i,j)\notin\eee \text{ and } j < i \rbrace}$. 
In~\cite{cordoba2020generating}, the authors suggest first triangulating the graph $\gggg$ to obtain a chordal graph, and then applying the Cholesky-based procedure from~\cite{cordoba2018metropolis}, as described above. The resulting matrix is the initialization of the partial orthogonalization algorithm.

In contrast, our proposed method (under some mild conditions) can generate correlation matrices with prescribed zero patterns, even for non-chordal graphs. Unlike the diagonal dominance approach, it does not suffer from excessively low correlation values. Compared to partial orthogonalization, our method is less sensitive to the initial matrix, in particular partial orthogonalization depends on the nodes numbering.

\section{Proposed approach} \label{sec:proposed}

The goal of our work is to generate correlation matrices in $\cc(\gggg)$, that is satisfying constraints \eqref{eq:constraint1}. Moreover, additional constraints can be added, depending on the context. One of our motivation is to construct correlation matrices with a distribution that resemble real-world data, particularly in neuroscience where the latter is shifted to positive values \cite{achard2022generation}. To reflect this property, we impose this additional constraint on the mean: for $b\geq -1$, \begin{equation}\label{eq:b}\frac{1}{2\vert \eee\vert}\sum_{i\neq j} c_{ij} \geq b.\end{equation} Taking $b\leq -1$ is equivalent to having no constraint. 

We seek to solve the following optimization problem:
\begin{equation} \label{eq:constraint2}
\begin{aligned}
& \underset{\cb}{\text{minimize}} & & \frac{1}{2}\lVert \cb - \bar{\cb} \rVert^2_F, \\
& \text{subject to} & &\text{constraints } \eqref{eq:constraint1} \text{ and } \eqref{eq:b},\
\end{aligned}
\end{equation}
with $\bar{\cb}$ a given arbitrary matrix. With real data, it can be the empirical correlation matrix. Note that solving~\eqref{eq:constraint2} ensures that the mean of the non-diagonal entries is at least $b$. An alternative approach would be to maximize $\text{tr}((\mathbf{1} - \mathbf{I})\mathbf{C})$, where $\mathbf{1}$ is a matrix of all ones and $\mathbf{I}$ is the identity matrix of the same dimension as $\mathbf{C}$, subject to the initial constraints~\eqref{eq:condition}. This formulation yields a unique solution that maximizes the mean of the non-diagonal entries. However, since our primary goal was to approximate the empirical correlation matrix as closely as possible (see Figure \ref{fig:correlations3}), we prioritize formulation~\eqref{eq:constraint2}.

The choice of the square of the Frobenius norm as an objective function is motivated by dealing with quadratic optimization, which often yields better convergence properties compared to linear objectives~\cite[Chapter 9]{boyd2004convex}.

If the objective function is convex and the intersection of the constraints forms a non-empty convex set, then the problem has a unique minimizer \cite{boyd2004convex}. Since the objective function in~\eqref{eq:constraint2} is convex, a solution exists whenever the constraints are feasible. Notably, the identity matrix satisfies the constraints~\eqref{eq:constraint1}, ensuring feasibility in the absence of the additional constraint~\eqref{eq:b}. In Section~\ref{sec:results}, we examine the impact of this additional constraint on solution feasibility.

\section{Results and discussions} \label{sec:results}
 In our simulations, we consider $\cb \in \mathbb{R}^{51\times51}$ and $\bar \cb$ a matrix of the same size whose entries are drawn from a uniform distribution over the interval $[-1, 1]$. For the pattern $\eee$ we use different random graph models, namely \emph{Erdős-Rényi}, \emph{Barabási-Albert}, \emph{Watts-Strogatz}, and \emph{Stochastic Block Model}. Additionally, we generate a chordal graph by triangulating a \emph{Barabási-Albert} graph. All graphs are generated using NetworkX~\cite{SciPyProceedings_11}. 
 
 We solved the optimization problem \eqref{eq:constraint2} in Python using the CVXPY library~\cite{cvxpy} and tested it on these graph models over 50 runs\footnote{Most of the computational analysis in this study was performed using the GRICAD infrastructure, supported by the Grenoble research community. The code to reproduce the experiments can be found in \cite{code}.}. In some cases, solving~\eqref{eq:constraint2} yields a matrix with a minimum eigenvalue close to zero while negative, which indicates that the matrix is not strictly PSD. To address this, we apply a shift and normalization strategy. Specifically, we add a small positive constant $\epsilon$ to the diagonal of the solution $\tilde{\mathbf{C}}$, i.e., $\tilde{\mathbf{C}_{\epsilon}} = \tilde{\mathbf{C}} + \epsilon \mathbf{I}$.  Subsequently, we normalize the shifted matrix $\tilde{\cb_{\epsilon}}$ along its main diagonal to obtain the desired correlation matrix $\mathbf{C}$, defined as $\mathbf{C} = \frac{1}{1+\epsilon}\tilde{\mathbf{C}_{\epsilon}}$. The matrix $\mathbf{C}$ is not the minimizer of the objective function, but it is a correlation matrix that satisfies the constraints\footnote{To be more precise, with this post-processing step the mean value changes, and then constraint \eqref{eq:b} may not be satisfied. Increasing $b$ to $b(1+\epsilon)$ allow to achieve our objective.} in \eqref{eq:constraint2}. %

\subsection{Comparison with other approaches} 

\begin{figure}[htbp]
	\centerline{\includegraphics[width=0.4\textwidth]{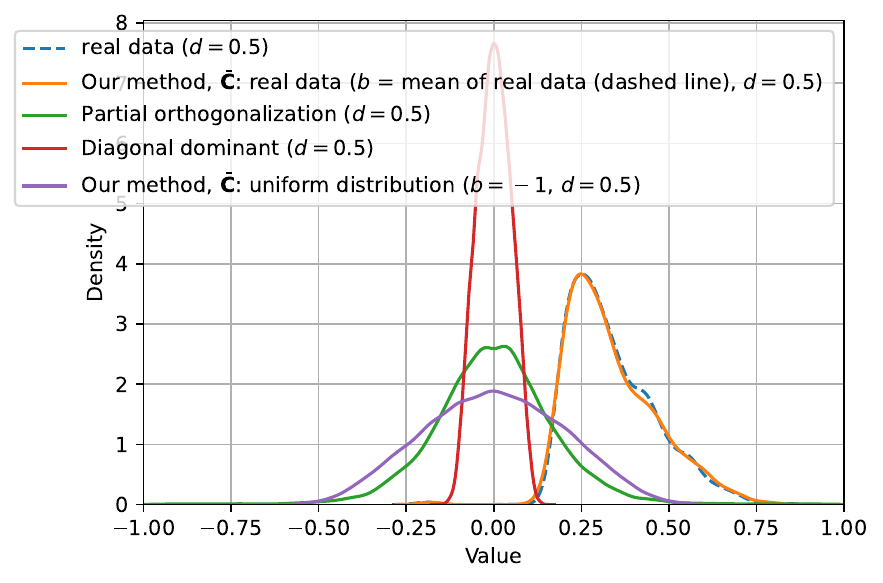}}
		\vspace*{-0.5\baselineskip}
	\caption{Density of non-diagonal, non-zero entries in generated correlation matrices using different methods (diagonal dominance, partial orthogonalization, and our method) compared to the correlation matrix obtained from rat fMRI data.}
	\label{fig:correlations3}  
	\vspace*{-0.5\baselineskip}	
\end{figure}

For comparison, we consider a graph with 51 nodes, i.e., $\mathbf{C} \in \mathbb{R}^{51 \times 51}$. Figure~\ref{fig:correlations3} shows the density of non-diagonal, non-zero entries. We compare our method with two other approaches: diagonal dominance and partial orthogonalization. Specifically, we generate 50 \emph{Erdős-Rényi} graphs using diagonal dominance, partial orthogonalization, and our proposed method. For our method, we set the target graph density to 0.5. The distribution of correlation values are represented by red, green, and purple lines, respectively ---the orange one is related to real data and explained below. We set the parameter  $b=-1$ in our algorithm to facilitate comparison with other algorithms that do not use a threshold. In our algorithm, both $\bar{\mathbf{C}}$ and the initial point for the diagonal dominance method are realizations of a uniform distribution, since we aim at generating random correlation matrices.  Notably, when using diagonal dominance, no positive perturbation is applied to ensure the matrix is PSD. As previously mentioned, the densities obtained using diagonal dominance are concentrated around low values (red line). Note that our approach (purple line) gives higher entries in the correlation matrix.

\subsection{Influence of the constraint on the mean}

We now examine the impact of the constraint in~\eqref{eq:b}, which modifies the centering of the distribution of correlation entries. In particular, it may be useful to control the signal-to-noise ratio in simulation studies, or to generate correlation matrices more alike real data. While Generative Adversarial Networks (GANs) can also generate correlation matrices similar to real data~\cite{marti}, they require a large dataset of observed correlation matrices, which may not always be available in practice.

In our context, we are motivated by a neuroscience application involving functional MRI data acquired on rats. The data are described and freely available~\cite{guillaume}. The recording duration is 30 minutes with a
repetition time of 0.5 seconds, and 3600 time points
are thus available at the end of the experiment. After the preprocessing explained in~\cite{guillaume}, time series of 51 brain regions for each rat were extracted. We then calculate the wavelet transform with the Daubechies wavelet of order 8 of the 51 signals. We are interested here in wavelet scale 4, corresponding to the frequency interval [0.06; 0.12] Hz.
There are then 122 wavelet coefficients available for each of the 51 regions. The distributions of the pairwise correlations between the wavelet coefficients of the regions for a given rat are presented in Figure~\ref{fig:correlations3}. 

In Figure~\ref{fig:correlations3}, the blue line represents the (empirical) correlation matrix of the rat fMRI data, while the orange line shows the distribution of the (theoretical) correlation matrix generated using our method. For the real data, we compute the graph density by selecting the 50\% of entries with the highest absolute values in the correlation matrix. In generating the synthetic matrix with our method, we set $d= 50\%$ and the parameter $b$ equal to the mean of the entries in the real-data correlation matrix corresponding to computed graph~($d=50\%$). The initial matrix $\bar{\cb}$ is here equal to the empirical correlation of the real data. Figure~\ref{fig:correlations3} shows that the distribution of the simulated data is close to the real one.

\begin{figure}[!ht]
	\centerline{\includegraphics[width=0.4\textwidth]{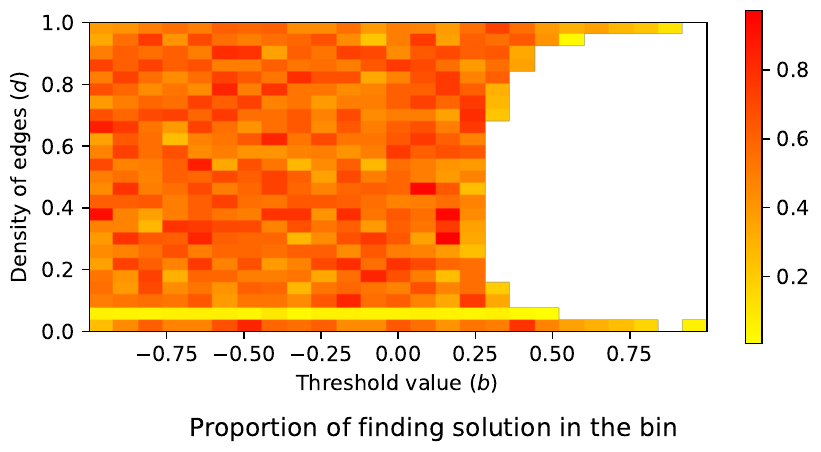}}
	\caption{Proportion of successfully finding a correlation matrix \mbox{$\cb \in \mathbb{R}^{51 \times 51}$} per bin over different threshold values of $b$, using an \emph{Erdős-Rényi} pattern. White areas indicate cases where no solution was found.}
	\label{fig:solution}  
	\vspace*{-0.5\baselineskip}	
\end{figure}

Adding constraint~\eqref{eq:b} may result in an optimization problem that has no solution. Figure~\ref{fig:solution}  illustrates the proportion of cases where a valid correlation matrix \mbox{$\cb \in \mathbb R ^{51 \times 51}$} is found for different graph densities and values of $b$ in~\eqref{eq:constraint2}. The figure considers \textit{Erdős-Rényi} graphs, but the results vary with different graph structures. In the figure, the face color is white if no solution is found, indicating that the constraints do not intersect in the optimization problem.

\subsection{Influence of the graph structure} 

\begin{figure}[htbp]
	\centerline{\includegraphics[width=0.32\textwidth]{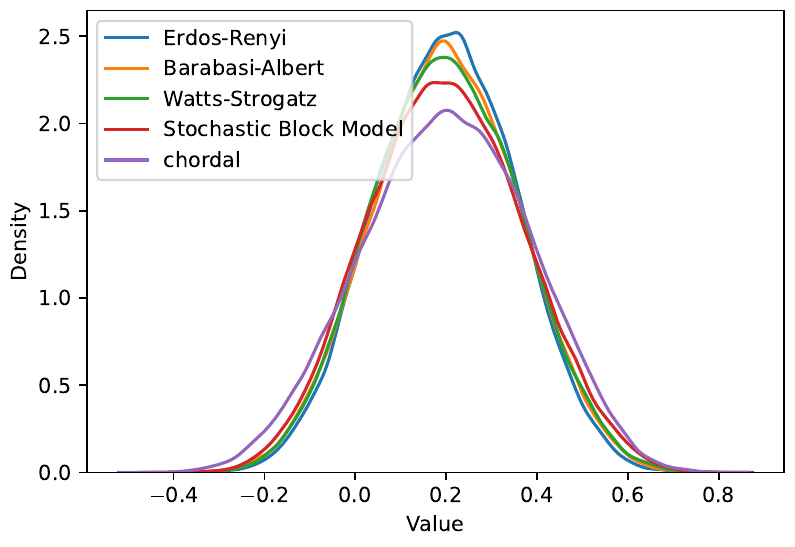}}
		\vspace*{-1\baselineskip}
	\caption{Density of non-zero, non-diagonal elements in the correlation matrix generated by our algorithm for different random graph models and the Chordal graph, given a graph edge density of \mbox{$d=0.5$} for \mbox{$\cb \in \mathbb{R}^{51 \times 51}$}. Results are obtained over 50 runs, with a threshold constraint of $b = 0.2$.}
	\label{fig:bd_2} 
\vspace*{-0.5\baselineskip}	
\end{figure}

This subsection explores how different graph structures influence the results. Figure~\ref{fig:bd_2} shows the distribution of correlation values for different $50\%$ dense graph types with $b = 0.2$. Correlation entries are centered around this value.

In Figure~\ref{fig:bd_2}, the type of graph has no significant effect on the density of non-zero and non-diagonal entries of the correlation matrices. Yet, as shown in \cite{achard2022generation}, we expect that the structure of the graph may have more influence when increasing the number of nodes.

\begin{figure}[htbp]
	\centerline{\includegraphics[width=0.4\textwidth]{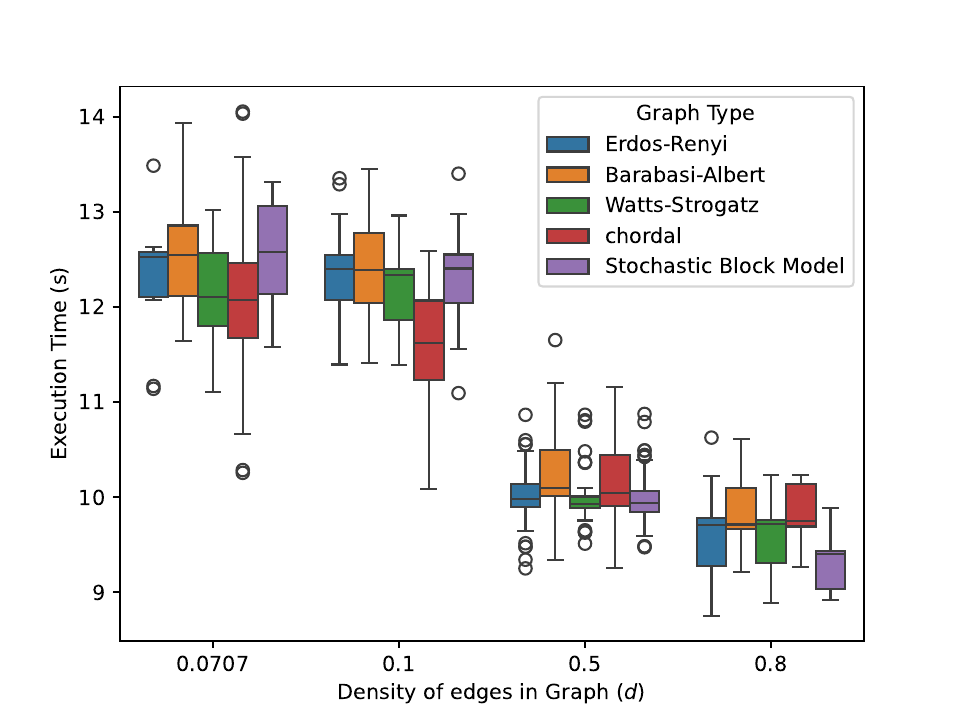}}
	\vspace*{-1\baselineskip}
	\caption{Execution time (in seconds) of our method for computing the correlation matrix \mbox{$\cb \in \mathbb{R}^{51 \times 51}$}, averaged over 50 runs for each box plot (representing a different graph type) as a function of the density of non-zero, non-diagonal elements in the correlation matrix. The parameter $b$ is set to~$-1$.}
	\label{fig:time}  
	\vspace*{-0.5\baselineskip}	
\end{figure}

The computational cost of solving the optimization problem is significantly higher than for methods like diagonal dominance or partial orthogonalization. While increasing the dimension of $\mathbf{C}$ generally raises the execution time, Figure~\ref{fig:time} compares execution times for a fixed dimension of 51 across varying graph densities and models. Overall, the execution time decreases as the graph density increases.

\section*{Conclusion and perspectives} \label{sec:conclusion}

In summary, our proposed method offers several advantages. It does not rely on a chordal structure, guarantees the generation of a positive semi-definite (PSD) matrix, and avoids generating near-zero entries in the resulting correlation matrix. Additionally, it enables the generation of a correlation matrix that corresponds to a graph related to an empirical correlation matrix. However, this approach comes at the cost of increased computational time.

As discussed, execution time increases with dimensionality. 
First we may study the capacity of the proposed method to handle with higher-dimensional settings. For instance, we will study how increasing the number of nodes affects the influence of graph structures.
For this purpose, the QSDPNAL algorithm~\cite{li2018qsdpnal} in MATLAB could be considered for solving quadratic objective functions in high-dimensional problems. Future work could explore alternative constraints, objective functions, and optimization algorithms to further enhance performance.

\bibliographystyle{plain}
\bibliography{biblo.bib}

\begin{thebibliography}{10}

\bibitem{abbe2018community}
Emmanuel Abbe.
\newblock Community detection and stochastic block models: recent developments.
\newblock {\em Journal of Machine Learning Research}, 18(177):1--86, 2018.

\bibitem{achard2022generation}
Sophie Achard, Ir{\`e}ne Gannaz, and K{\'e}vin Polisano.
\newblock G{\'e}n{\'e}ration de mod{\`e}les graphiques.
\newblock In {\em GRETSI 2022-XXVIII{\`e}me Colloque francophone de traitement
  du signal et des images}, pages 1--3, 2022.

\bibitem{akbani2014pan}
Rehan Akbani, Patrick Kwok~Shing Ng, Henrica~MJ Werner, Maria Shahmoradgoli,
  Fan Zhang, Zhenlin Ju, Wenbin Liu, Ji-Yeon Yang, Kosuke Yoshihara, Jun Li,
  et~al.
\newblock A pan-cancer proteomic perspective on the {C}ancer {G}enome {A}tlas.
\newblock {\em Nature communications}, 5(1):3887, 2014.

\bibitem{albert1999diameter}
R{\'e}ka Albert, Hawoong Jeong, and Albert-L{\'a}szl{\'o} Barab{\'a}si.
\newblock Diameter of the world-wide web.
\newblock {\em Nature}, 401(6749):130--131, 1999.

\bibitem{armour2017network}
Cherie Armour, Eiko~I Fried, Marie~K Deserno, Jack Tsai, and Robert~H Pietrzak.
\newblock A network analysis of {DSM-5} posttraumatic stress disorder symptoms
  and correlates in {US} military veterans.
\newblock {\em Journal of anxiety disorders}, 45:49--59, 2017.

\bibitem{guillaume}
Guillaume J-PC Becq, Tarik Habet, Nora Collomb, Margaux Faucher, Chantal
  Delon-Martin, V{\'e}ronique Coizet, Sophie Achard, and Emmanuel~L Barbier.
\newblock Functional connectivity is preserved but reorganized across several
  anesthetic regimes.
\newblock {\em NeuroImage}, 219:116945, 2020.

\bibitem{boyd2004convex}
Stephen Boyd and Lieven Vandenberghe.
\newblock {\em Convex optimization}.
\newblock Cambridge university press, 2004.

\bibitem{chib1995understanding}
Siddhartha Chib and Edward Greenberg.
\newblock Understanding the {M}etropolis-{H}astings algorithm.
\newblock {\em The American Statistician}, 49(4):327--335, 1995.

\bibitem{cordoba2018metropolis}
Irene C{\'o}rdoba, Gherardo Varando, Concha Bielza, and Pedro Larra{\~n}aga.
\newblock A fast metropolis-hastings method for generating random correlation
  matrices.
\newblock In {\em International Conference on Intelligent Data Engineering and
  Automated Learning}, pages 117--124. Springer, 2018.

\bibitem{cordoba2020generating}
Irene C{\'o}rdoba, Gherardo Varando, Concha Bielza, and Pedro Larra{\~n}aga.
\newblock On generating random {G}aussian graphical models.
\newblock {\em International Journal of Approximate Reasoning}, 125:240--250,
  2020.

\bibitem{cordoba2020unifying}
Irene Córdoba.
\newblock {\em Unifying methodologies for graphical models with {G}aussian
  parameterization}.
\newblock PhD thesis, Universidad Politécnica de Madrid, 2020.

\bibitem{cvxpy}
Steven Diamond and Stephen Boyd.
\newblock {CVXPY}: A {P}ython-embedded modeling language for convex
  optimization.
\newblock {\em Journal of Machine Learning Research}, 17(83):1--5, 2016.

\bibitem{engel2017overview}
Jasper Engel, Lutgarde Buydens, and Lionel Blanchet.
\newblock An overview of large-dimensional covariance and precision matrix
  estimators with applications in chemometrics.
\newblock {\em Journal of Chemometrics}, 31(4):e2880, 2017.

\bibitem{erdos1960evolution}
Paul Erdos and Alfr{\'e}d R{\'e}nyi.
\newblock On the evolution of random graphs.
\newblock {\em Publ. math. inst. hung. acad. sci}, 5(1):17--60, 1960.

\bibitem{code}
Ali Fahkar, Polisano Kévin, Irène Gannaz, and Sophie Achard.
\newblock Code implementation of generating correlation matrices with graph
  structures using convex optimization.
\newblock Gricad-gitlab repository, 2025.
\newblock
  https://gricad-gitlab.univ-grenoble-alpes.fr/polisank/generating-correlation-matrices-with-graph-structures-using-convex-optimization.

\bibitem{Giraud}
Christophe Giraud.
\newblock {\em Introduction to high-dimensional statistics}.
\newblock CRC Press, 2021.

\bibitem{grechkin2015pathway}
Maxim Grechkin, Maryam Fazel, Daniela Witten, and Su-In Lee.
\newblock Pathway graphical lasso.
\newblock In {\em Proceedings of the AAAI conference on artificial
  intelligence}, volume~29, 2015.

\bibitem{SciPyProceedings_11}
Aric~A. Hagberg, Daniel~A. Schult, and Pieter~J. Swart.
\newblock Exploring network structure, dynamics, and function using networkx.
\newblock In Ga\"el Varoquaux, Travis Vaught, and Jarrod Millman, editors, {\em
  Proceedings of the 7th Python in Science Conference}, pages 11 -- 15,
  Pasadena, CA USA, 2008.

\bibitem{horn2012matrix}
Roger~A Horn and Charles~R Johnson.
\newblock {\em Matrix analysis}.
\newblock Cambridge University Press, 2012.

\bibitem{huang2010learning}
Shuai Huang, Jing Li, Liang Sun, Jieping Ye, Adam Fleisher, Teresa Wu, Kewei
  Chen, Eric Reiman, and Alzheimer's Disease~NeuroImaging Initiative.
\newblock Learning brain connectivity of {A}lzheimer's disease by sparse
  inverse covariance estimation.
\newblock {\em NeuroImage}, 50(3):935--949, 2010.

\bibitem{hull2012risk}
John Hull.
\newblock {\em Risk management and financial institutions,+ Web Site}, volume
  733.
\newblock John Wiley \& Sons, 2012.

\bibitem{joe2006generating}
Harry Joe.
\newblock Generating random correlation matrices based on partial correlations.
\newblock {\em Journal of Multivariate Analysis}, 97(10):2177--2189, 2006.

\bibitem{kimmig2015lifted}
Angelika Kimmig, Lilyana Mihalkova, and Lise Getoor.
\newblock Lifted graphical models: a survey.
\newblock {\em Machine Learning}, 99:1--45, 2015.

\bibitem{koller2009probabilistic}
Daphne Koller and Nir Friedman.
\newblock {\em Probabilistic graphical models: principles and techniques}.
\newblock MIT press, 2009.

\bibitem{lewandowski2009generating}
Daniel Lewandowski, Dorota Kurowicka, and Harry Joe.
\newblock Generating random correlation matrices based on vines and extended
  onion method.
\newblock {\em Journal of multivariate analysis}, 100(9):1989--2001, 2009.

\bibitem{li2018qsdpnal}
Xudong Li, Defeng Sun, and Kim-Chuan Toh.
\newblock Qsdpnal: A two-phase augmented lagrangian method for convex quadratic
  semidefinite programming.
\newblock {\em Mathematical Programming Computation}, 10:703--743, 2018.

\bibitem{marti}
Gautier Marti, Victor Goubet, and Frank Nielsen.
\newblock ccorrgan: Conditional correlation gan for learning empirical
  conditional distributions in the elliptope.
\newblock In {\em International Conference on Geometric Science of
  Information}, pages 613--620. Springer, 2021.

\bibitem{pearl2014probabilistic}
Judea Pearl.
\newblock {\em Probabilistic reasoning in intelligent systems: networks of
  plausible inference}.
\newblock Elsevier, 2014.

\bibitem{pourahmadi2015distribution}
Mohsen Pourahmadi and Xiao Wang.
\newblock Distribution of random correlation matrices: {H}yperspherical
  parameterization of the {C}holesky factor.
\newblock {\em Statistics \& Probability Letters}, 106:5--12, 2015.

\bibitem{salas1999gershgorin}
Hector~N Salas.
\newblock Gershgorin's theorem for matrices of operators.
\newblock {\em Linear algebra and its applications}, 291(1-3):15--36, 1999.

\bibitem{vandenberghe2015chordal}
Lieven Vandenberghe and Martin~S Andersen.
\newblock Chordal graphs and semidefinite optimization.
\newblock {\em Foundations and Trends{\textregistered} in Optimization},
  1(4):241--433, 2015.

\bibitem{watts1998collective}
Duncan~J Watts and Steven~H Strogatz.
\newblock Collective dynamics of ‘small-world’networks.
\newblock {\em nature}, 393(6684):440--442, 1998.

\end{thebibliography}

\end{document}